\documentclass[twocolumn, trackchanges]{aastex63}
\pdfoutput=1
\usepackage{mathtools}
\usepackage[utf8x]{inputenc}

\received{2020 December 7}
\revised{2021 May 9}
\accepted{2021 May 22}

\graphicspath{{./}{figures/}}

\begin{document}

\title{The Effect of the Angular Momentum in the Formation and Evolution of Low Surface Brightness Galaxies}

\correspondingauthor{Vicente Salinas}
\email{vhsalinas@uc.cl}

\author[0000-0002-6858-4976]{Vicente H. Salinas}
\affiliation{Instituto de Astrofísica, Pontificia Universidad Católica de Chile}

\author[0000-0002-8835-0739]{Gaspar Galaz}
\affiliation{Instituto de Astrofísica, Pontificia Universidad Católica de Chile}
\email{ggalaz@astro.puc.cl}

\begin{abstract}

Using observed data from the literature, we compare in one single publication the angular momentum (AM) of low surface brightness galaxies (LSBGs), with that of high surface brightness galaxies (HSBGs), a comparison that either is currently spread across many unconnected references, or simply does not exist. Partly because of the subject, this has received little attention outside the realm of simulations. We use previous results of the stellar specific AM $j_{*}$ from the SPARC database containing Spitzer $3.6$ $\mu m$ photometry and accurate H\,\textsc{i} rotation curves from \citeauthor{2016AJ....152..157L} using a sample of 38 LSBGs and 82 HSBGs. We do this with the objective of comparing both galaxy populations, finding that LSBGs are higher in the Fall relation by about $0.174$ dex. Aditionally, we apply and test different masses and formation models to estimate the spin parameter $\lambda$, which quantifies the rotation obtained from the tidal torque theory, finding no clear evidence of a difference in the spin of LSBGs and HSBGs under a classic disk formation model that assumes that the ratio ($f_{j}$) between $j_{*}$ and the specific AM of the halo is $\sim 1$. In another respect, by using the biased collapse model, where $f_{j}$ depends on the star formation efficiency, it was found that LSBGs clearly show higher spin values, having an average spin of $\sim 2$ times the average spin of HSBGs. This latter result is consistent with those obtained from simulations by \citeauthor{Dalcanton_1997}

\end{abstract}


\section{Background and Introduction to the Problem}
When we look at the properties of galaxies, the angular momentum (AM hereafter) $J$, along with the mass $M$ and energy $E$, emerges as one of the most important and studied physical parameters. The relationship between the AM with the evolution and formation of galaxies has been studied in various ways. Examples include studies of the origin of the AM in galaxies \citep[e.g.,][]{1969ApJ...155..393P}, the conservation of AM \citep[e.g.,][]{Fall_1983}, the relationship with morphological types \citep[e.g.,][]{2012ApJS..203...17R}, the effect of mergers in the gain or loss of AM \citep[e.g.,][]{Lagos_2017}, the many numerical simulations that try to match the observations \citep[e.g.,][]{Lagos_2017,2020MNRAS.496.3996K,2000ApJ...538..477N}, and others. In particular, there is an aspect that, although it has been studied, still holds many loose ends, and that is the case of the AM of low surface brightness galaxies (LSBGs; e.g., \citeauthor{1996MNRAS.283...18D} \citeyear{1996MNRAS.283...18D}; \citeauthor{Schombert_2001} \citeyear{Schombert_2001}; \citeauthor{ONeil_2007} \citeyear{ONeil_2007}; \citeauthor{Galaz_2015} \citeyear{Galaz_2015}), and how it compares with that of high surface brightness galaxies (HSBGs). The very existence of LSBGs posses a challenge to modern galaxy formation and evolution theories. \citet{Schombert_2001} compares the study of LSBGs to psychologist studying extreme behaviours. Like with people, if there is a hope to understand galaxies, a need to be able to explain the extreme conditions that LSBGs present is required. Furthermore, studies indicate that LSBGs dominate the volume density of galaxies \citep[e.g.,][]{Dalcanton_1997,O_Neil_2000}. It seems that, in reality, these extreme behaviors are quite common. In another respect, LSBGs actually provide great opportunities to understand galaxy evolution, as they are considered to be dominated by dark matter (DM), which makes them useful to test DM theories. And because they are believed to evolve slowly compared to HSBGs, they are sometimes referred as fossils of galaxy formation. 

Following this logic, if the aim is to understand the AM of galaxies, it might be insightful to take a look at the AM of LSBGs. For example, a major issue regarding the AM of galaxies is the need to understand the reasons for the observed differences in AM of spirals and elliptical galaxies, with observations that result in spirals having systematically higher AM. This issue began from observational results, but on the other hand, in the LSBGs versus HSBGs case, this comparison arose mainly from simulations \citep[e.g.,][]{Dalcanton_1997}, resulting in LSBGs having, in general, higher AM. Nevertheless, few studies had actually put together both observed measurements of the AM of LSBGs and HSBGs, side by side, and contrasting results with models. With this work the goal is not to provide new AM computations of galaxies, as those already exists in the literature. Instead, we aim to connect the existing results and present them under the context of an LSBGs versus HSBGs comparison.

In the context of the Cold Dark Matter model, galaxies are formed in the center of DM halos, and the origin of the AM is explained by the so-called Tidal Torque Theory (TTT), first introduced by \citet{Hoyle_1949} and later expanded upon by \cite{1969ApJ...155..393P} in the gravitational instability picture before the gravitational collapse. Halos acquire their AM via tidal torques exerted by neighboring overdensities of other halos, and then DM and baryons start to collapse in the overdense regions conserving AM. Since DM does not dissipate energy, its collapse halts when the system virializes. We should add that the gas, which is also being affected by the tidal torques, can dissipate energy. Because of this, the gas can lose potential energy and fall to the center of the halo, where it becomes cold and dense enough for nuclear fusion to ignite and thus starts star formation.  

\begin{figure}[ht!]
\plotone{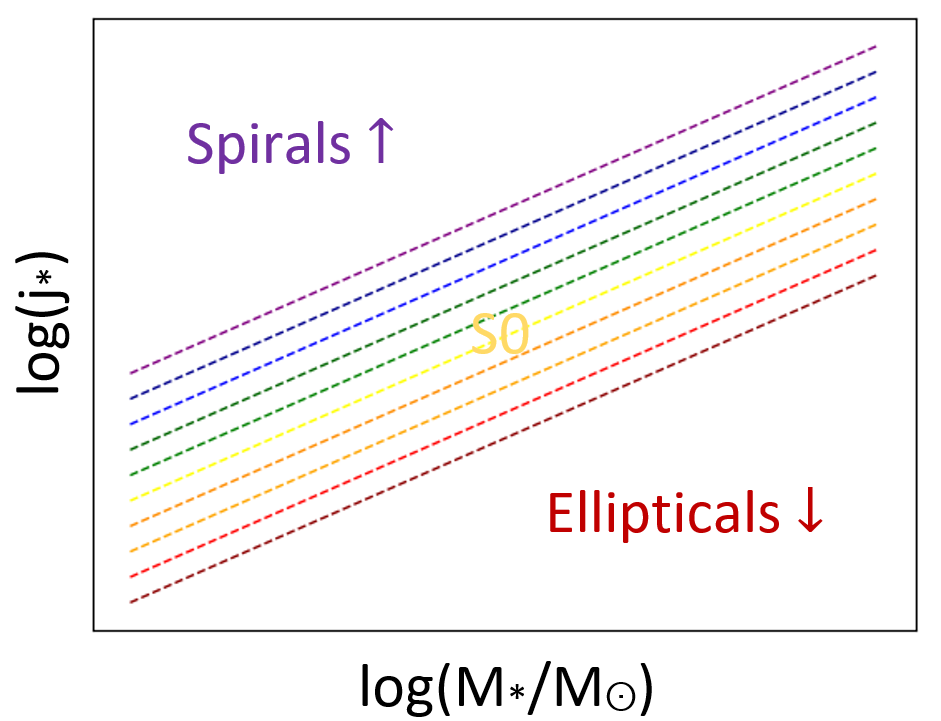}
\caption{Schematic of the Fall Relation \citep{Fall_1983}. On top disk-like galaxies appear (Sc,Sb,Sa), and at the bottom bulge-like galaxies are situated (ellipticals), with lenticular galaxies in between spirals and ellipticals. \label{fig:EF}}
\end{figure}

The AM obtained in this way can be quantified in the form of the dimensionless spin parameter $\lambda=J|E|^{1/2}G^{-1}M^{-5/2}$ \citep{Peebles_1971}, with $G$ being the gravitational constant. Or, alternatively, some authors use \citep{Bullock_2001}

\begin{equation}
\lambda'=\frac{j}{\sqrt{2}R_\textnormal{vir}V_\textnormal{vir}},
\label{eq:spin}
\end{equation}

where $R_\textnormal{vir}$ is the virial radius, $V_\textnormal{vir}=\sqrt{GM_\textnormal{vir}/R_\textnormal{vir}}$ is the circular velocity at $R_\textnormal{vir}$, with $M_\textnormal{vir}$ being the virial mass, and $j$ is the specific AM, defined by $j\equiv J/M$. Instead of using $J$, we define the specific AM $j$, because it contains information about both the length scale, $R_{d}$, and rotational velocity. This value, however, depends on the mass of the galaxy since the rotational velocity is correlated with mass by the Tully-Fisher relation \citep{1977A&A....54..661T}. This is the advantage the spin parameter has over the specific AM, since the spin does not depend on the mass. But, in another respect, $j$ is easier to estimate (needing less assumptions than the spin), and can still provide much insight about the formation of galaxies, especially from the $j-M$ diagram.

A classic example of the specific AM being capable of explaining the possible evolution of galaxies, is the case of the Fall relation, which shows that galaxies of different morphological types fall in different regions of the $j_{*}-M_{*}$ diagram (the $*$ subscript refers to the stellar component). It was first introduced by \citet{Fall_1983}, who shows that spiral galaxies follow a tight trend, while ellipticals are more spread out and located bellow the spirals in the diagram. Later, \cite{2012ApJS..203...17R} and \cite{Fall_2013} reproduced the same diagram with more data and additional techniques, showing that at a fixed mass $M_{*}$, elliptic galaxies have about $\sim4$ times less AM than spirals. They claimed that galaxies of different morphological types would fall at different regions in the diagram, so that they would have trends with the same slope of $\sim2/3$ but different zero-point, with the most disk-like galaxies on top and the more bulge-like on the bottom of the diagram, as shown in Figure \ref{fig:EF}. They obtain results consistent with this model, and find that, on average, lenticular galaxies fill the gap between ellipticals and spirals, suggesting that the Hubble sequence could be replaced by a more physically motivated classification, based on angular momentum and stellar mass.

It is rather apparent that the DM must have a big effect on the AM distribution of baryons, considering that the strongest evidence supporting the existence of DM halos is the rotation curves of galaxies \citep{Rubin_1970}, which are a key ingredient of AM. Not only the rotational velocity, but also the radial distribution of baryons would be affected by the DM (more on that later in this paper). Furthermore, a more straightforward connection can be made, since from the TTT we can arrive at the relation 

\begin{equation}
j_{h} \propto \lambda M_{h}^{2/3},
\label{eq:jh}
\end{equation}

with $j_{h}$ being the specific AM of the halo, and $M_{h}$ being the mass of the halo. The $j_{*}-M_{*}$ relation actually provides an important connection with the DM component of the AM. If the retained AM is defined  as $f_{j}=j_{*}/j_{h}$, the star formation efficiency as $f_{*}=M_{*}/M_{h}$, and the relation (\ref{eq:jh}) is multiplied by $f_{j}$ and $f_{*}^{-2/3}$ the following expression is obtained,

\begin{equation}
j_{*} \propto \lambda f_{j} f_{*}^{-2/3} M_{*}^{2/3}.
\label{eq:js}
\end{equation}

This means that slopes of $\sim 2/3$ in the Fall relation imply that the factor $\lambda f_{j} f_{*}^{-2/3}$ is roughly constant, and any deviations from this slope would suggest systematic changes in this physically related factor.

Aside from the TTT, another mechanism in which galaxies might gain or lose specific AM is the merging process \citep[e.g.,][]{Lagos_2017}. Since we know from \cite{1972ApJ...178..623T} that a massive elliptical could be formed after a major merger of two disk galaxies, this could explain the position of the ellipticals in the diagram. Fortunately, for the purpose of this work, there is not a strong reason to consider mergers when comparing LSBGs with HSBGs, since LSBGs seem to be mostly isolated systems  \citep{2004A&A...422L...5R,tanoglidis2020shadows}, with very late Hubble types.

LSBGs are commonly defined as galaxies with central surface brightness (SB) $\mu_{0}$ fainter than $22$ B mag arsec$^{-2}$, which corresponds to a value outside the range that \citet{Freeman_1970} initially find for the central SB of spirals and S0, of $\mu_{0}=21.65\pm0.30$ B-mag arsec$^{-2}$. Initially, it was thought that galaxies with fainter values do not exist, but later \citet{Disney_1976} showed that this apparent physical limit (called initially the ``Freeman limit") was indeed a bias from the photographic plates: the value was basically the sky brightness registered by the plates themselves. Any disk fainter than this value would be submerged in the sky brightness. Disney showed then the existence of a large fraction of LSBGs, arguing that they were not visible because of the difficulty of detecting galaxies of faint SB given the limitation of photographic plates. Perhaps the first grand design LSB galaxy that really surprised the astronomical community was the giant LSB Malin 1, discovered in 1986 \citep{Bothun_1987}, and since then many others LSBGs have been found, giving birth to a new area of research. 

One recursive question is whether LSBGs follow the same Tully-Fisher (TF) relation as normal galaxies \citep[e.g.][]{10.1093/mnras/273.1.L35,Chung_2002}. At present we found that they mostly do follow the same relation, and the implications of this are best shown by \citet{10.1093/mnras/273.1.L35} in a small calculation; since by definition LSBGs have fainter $I_{0}$ than HSBGs, where $I_{0}$ denotes the central SB in physical units, the difference in $I_{0}$ must be compensated by a difference in mass to light ratio $M/L$. Consequently, from the relations $M \propto v_\textnormal{max}^{2}R_{d}$ and $L \propto I_{0}R_{d}^{2}$, where $L$ is luminosity, and $v_\textnormal{max}$ is the maximum rotation velocity, we can write

\begin{equation}
L\propto \frac{v_\textnormal{max}^{4}}{I_{0}(M/L)^{2}}.
\end{equation}

Since the TF relation expresses that $L\propto v_\textnormal{max}^{4}$, then $I_{0}(M/L)^{2}$ needs to remain constant, implying higher $M/L$ ratios for LSBGs. Also, from these very same relations we have  $(M/L)\propto (R_{d}^{2}/M)=\bar{\sigma}^{-1}$, with $\bar{\sigma}$ the mean surface mass density. Thus, $I_{0}\propto{\sigma}^{2}$, proving that LSBGs are less dense than HSBGs. On the one hand, both of these implications indicate that LSBGs have a higher fraction of gas and DM, and lower star formation rates. On the other hand, simulations made by \citet{Dalcanton_1997} show that rotation curves with high AM have higher $M/L$ ratios due to the dominance of DM at any radius, suggesting that high AM systems should be found mostly in LSBGs. They also find that changes in $\lambda$ affect the form of the rotation curves: when $\lambda$ is increased, the collapse factor of baryons decreases, leaving them at a higher radius, so that the baryonic fraction at any radius is smaller. The latter shows that the rotation curve of LSBGs is mostly dominated by the DM distribution, with curves that rise more slowly, rather than steeply.

Beside simulations, an important question is what are the observations telling us about the spin of LSBGs? The problem is that there are few direct computations of the spin from observed data in LSBGs and HSBGs, because of the trouble in estimating the AM of DM. Another problem is that works about the observed AM in galaxies often have other main interests, which have distributed information regarding the AM in LSBGs as minor results in various unconnected papers. Thus, with this work the purpose is, on the one hand, to compute the spin distributions of LSBGs and HSBGs from observational data using a variety of different common assumptions, and on the other hand, to take those already existing results about AM in galaxies and present them in a single publication, under a direct LSBG versus HSBG comparison. This paper should provide a simple but useful connecting point that organizes what we know about the AM of LSBGs.

\section{Data} 

\subsection{Data and sample}

The data used in this paper comes from the Spitzer Photometry and Accurate Rotation Curves (SPARC) database \citep[for a complete description of the SPARC sample, see][]{2016AJ....152..157L}, which we chose because it provides a complete, medium-sized sample with enough information to compare and study the AM of the two galaxy populations in question. It is comprised of a collection of 175 H\,\textsc{i} rotation curves from various compilations made with the WSRT, VLA, ATCA, and GMRT telescopes, along with 3.6 $\mu m$ near-infrared images from the Spitzer telescope. SPARC aims to have a broad galaxy sample, with morphologies ranging from S0 to Im/BCD, luminosity values from $\sim 10^{7}$ to $\sim 10^{12}$ $L_{\odot}$, effective SB from $\sim 5$ to $\sim 5000$ $L_{\odot}$ pc$^{-2}$, and rotational velocities ranging from $\sim 20$ to $\sim300$ km s$^{-1}$. Galaxy distances were measured in three different ways: accurate distances from the red giant branch, cepheids, and supernovae (errors between $\sim 5 \%$ and $\sim 10 \%$), distances from the Ursa Major cluster, and uncertainties estimated from the Hubble flow assuming $H_{0}=73$ km s$^{-1}$ Mpc$^{-1}$, with errors between $\sim10\%$ and $\sim30\%$ (these errors account for peculiar velocities and an uncertainty of $7\%$ in $H_{0}$). They perform surface photometry at 3.6 $\mu m$ and obtain  central SB $\mu_{0}$ for galaxies by fitting exponential functions to the outer parts of the SB profiles. They also estimate the disk scale length $R_{d}$.

\begin{figure}[ht!]
\plotone{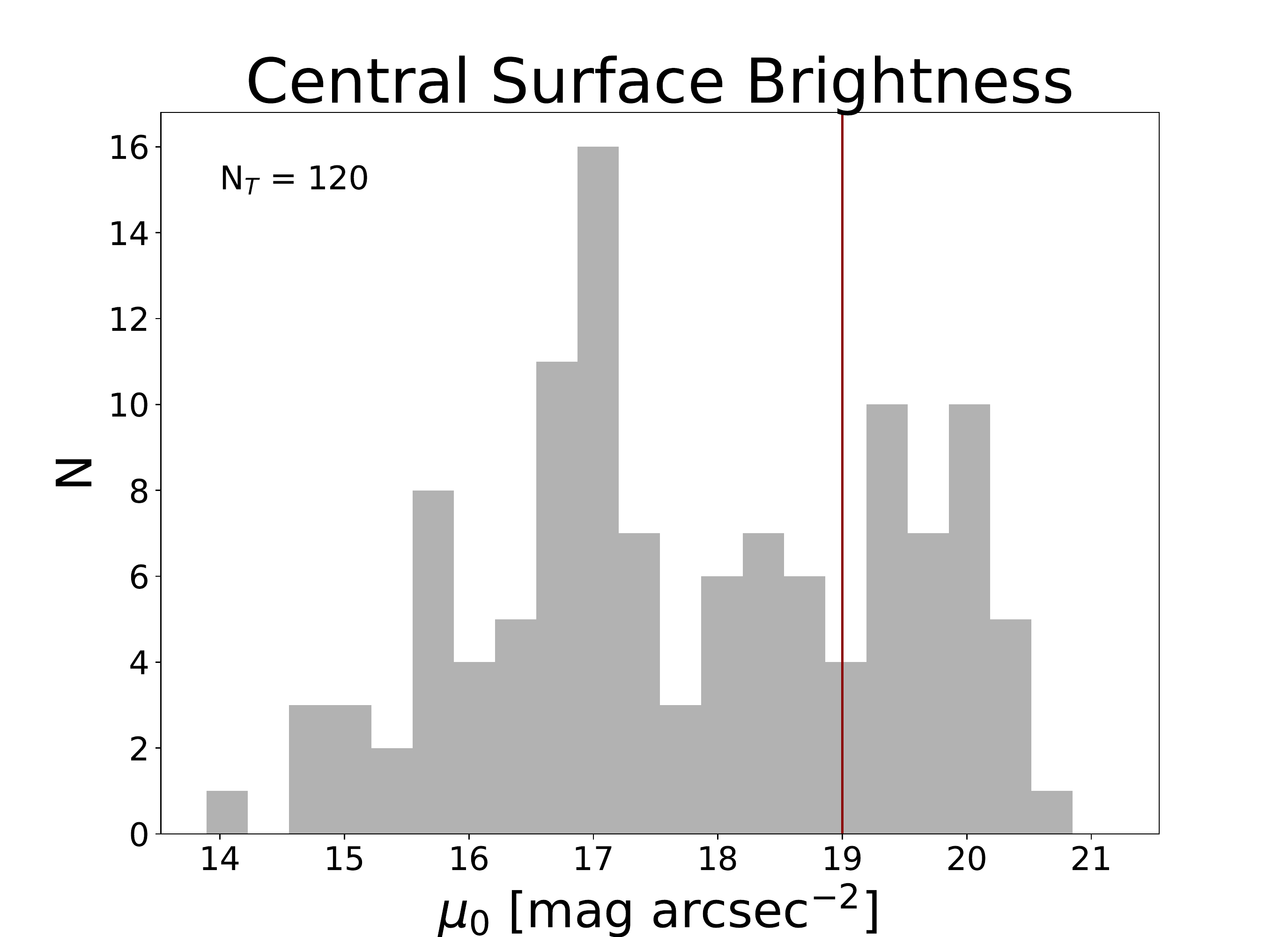}
\caption{Central SB (3.6 $\mu m$) histogram of the final sample. The red vertical line marks the separation between HSBGs and LSBGs. 82 are HSBGs and 38 are LSBGs. \label{fig:muH}}
\end{figure}

The selected sample for this work is similar to most of the existing SPARC samples used in the literature. Using a quality flag to select galaxies, only flags of less than 3 would be used. Only galaxies with inclinations of $30^{o}\leq i$ were considered, due to the uncertainty in their rotation velocity. And Finally, only galaxies with valid measurements of the mean velocity along the flat part of the rotation curve, $v_{f}$, were included. This leaves a total sample of 120 galaxies. Subsequently, and in order to clearly separate LSBGs from HSBGs, the central SB, $I_{0}$, in physical units ($L_{\odot}$ pc$^{-2}$), was converted to magnitude units by the formula\footnote{Borrowed from \cite{Schulz_2017}.}

\begin{equation}
\mu_{0} = m_{\odot,3.6\mu m} + 21.572-2.5log(I_{0})
\end{equation}

where $m_{\odot,3.6\mu m}=3.26$ is the magnitude of the Sun at 3.6 $\mu m$ \citep{2018ApJS..236...47W}. All galaxies with $\mu_{0}>19$  ($3.6$ $\mu m$) mag arsec$^{-2}$ were considered as LSBGs, as it is the limit \citet{2014PASA...31...11S} find in a sample of LSBGs, from which most of the LSBGs in SPARC were taken. The total central SB distribution is shown in the histogram of Figure \ref{fig:muH}.

\begin{figure}[ht!]
\plotone{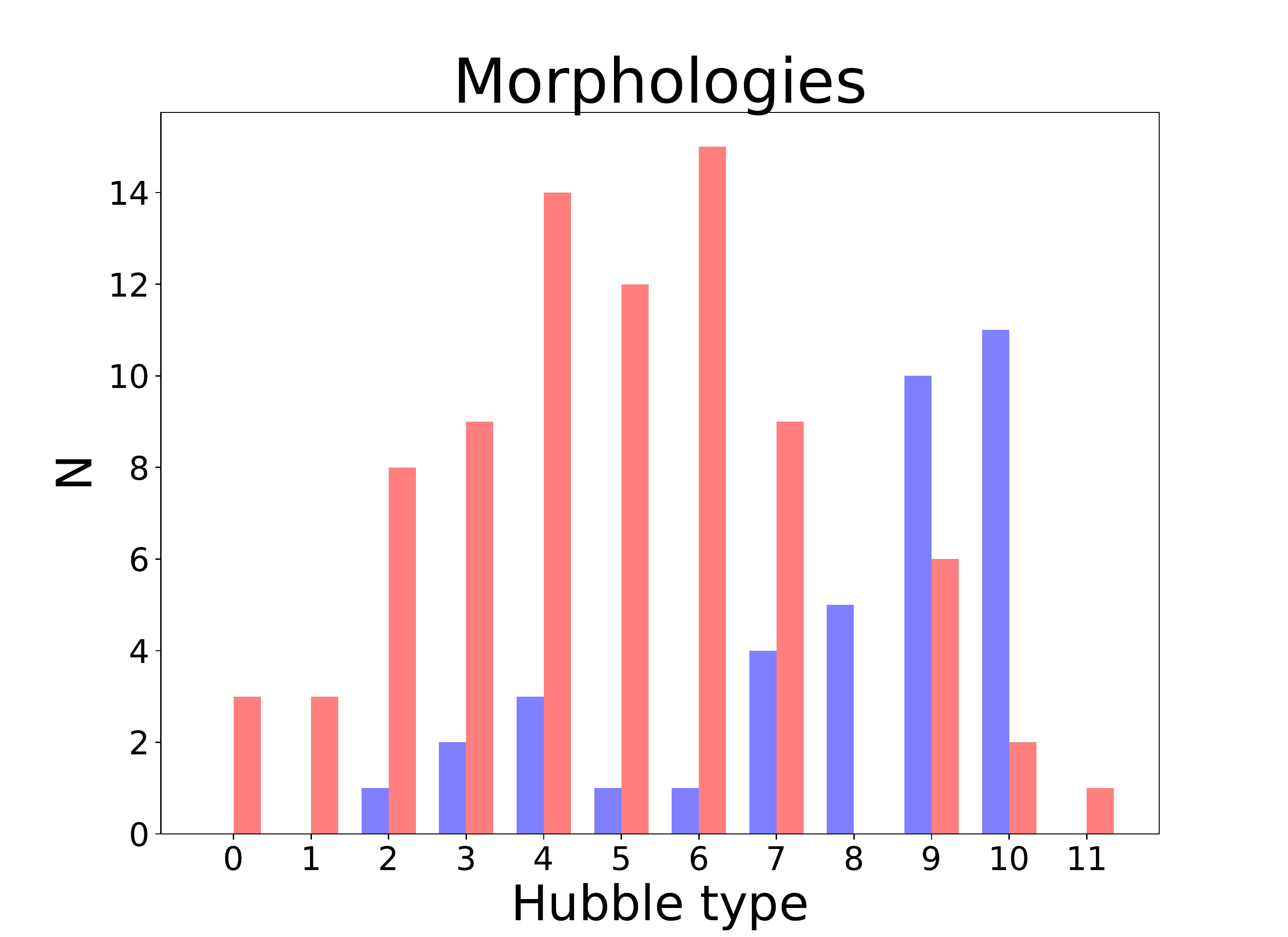}
\caption{Morphology distribution of LSBGs (blue) and HSBGs (red). The Hubble types are given in numerical form such that 0=S0, 1 = Sa, 2=Sab, 3=Sb, 4=Sbc, 5=Sc, 6=Scd, 7=Sd, 8=Sdm, 9=Sm, 10=Im, 11=BCD. The morphological classification is provided by \citet{2016AJ....152..157L}.\label{fig:morph}}
\end{figure}

Morphologies of the resulting sample are presented in Figure \ref{fig:morph}, which are also provided by the SPARC database \citep{2016AJ....152..157L}. From here it can be noted that the LSBGs are mostly classified as very late Hubble types, such as Sm and Im. This is because Hubble classification does not represent very well the physical differences in LSBGs \citep{1995AJ....109.2019M}, and tends to classify LSBGs as very late types.

\subsection{Computing the stellar angular momentum}

The AM of the $i$th component (stars, gas, ...) in a volume $V$ is given by 

\begin{equation}
J_{i}=\iiint_V v_{i}(\mathbf{r}) \Sigma_{i}(\mathbf{r}) \,dV,
\end{equation}

where $v_{i}(\mathbf{r})$ is the rotation velocity at the point $\mathbf{r}$, and $\Sigma_{i}(\mathbf{r})$ is the mass density at $\mathbf{r}$. In the case of a disk, this translates into

\begin{equation}
J_{i}=2\pi \int_{0}^{R_\textnormal{max}} v_{i}(r) \Sigma_{i}(r)r^{2} dr,
\label{eq:mom}
\end{equation} 

where $r$ is the galactocentric radius and $R_\textnormal{max}$ is the maximum radius of the disk. Similarly, the mass for a disk is

\begin{equation}
M_{i}=2\pi \int_{0}^{R_\textnormal{max}} \Sigma_{i}(r) r dr.
\label{eq:mass}
\end{equation}

Subsequently, the specific AM can be calculated using Equations (\ref{eq:mom}) and (\ref{eq:mass}).

In this way, if the rotation velocity is measured, and the mass density radial profile is computed, $j$ can be also computed. In practice, however, this is not always possible. To estimate the specific AM without measuring $\Sigma_{i}(r)$, many authors use the relation $j_{*} \approx 2R_{d}v_{f}$ for the stellar component of disk galaxies. This is the method that, for example, \citet{2012ApJS..203...17R} use in their paper for the disk-like galaxies. The simplicity of this relation makes it convenient for large data and quick estimates, although it is not as accurate as computing the AM through the density profile. Nevertheless, for the SPARC data the radial profiles are possible to compute by assuming a constant stellar mass-to-light ratio. Multiple studies suggest that the stellar mass-to-light ratios in the near-infrared are constant for a large range of morphology types and masses \citep[e.g.,][]{2014PASA...31...36S}. Hence, we use a similar procedure as in \cite{2018A&A...612L...6P}. Here, we consider a bulge/disk  decomposition \citep[from ][]{2017ApJ...836..152L}, where a stellar mass-to-light ratio of $\Upsilon_{b}=0.7$ $M_{\odot}/L_{\odot}$ is used for the bulge, and a stellar mass-to-light ratio of $\Upsilon_{d}=0.5$ $M_{\odot}/L_{\odot}$ is used for the disk. Resulting in the radial profile being described by

\begin{equation}
\Sigma_{*}(r)= I_{b}(r)\Upsilon_{b} + I_{d}(r)\Upsilon_{d},
\label{eq:sigma}
\end{equation}

where $I_{b}$ and $I_{d}$ are the SB of the the bulge and disk, respectively. Then the stellar AM is calculated using Equations (\ref{eq:sigma}) and (\ref{eq:mom}), together with the values of $v_{*}(r)$ from the rotation curves, with the mass calculated using Equations (\ref{eq:sigma}) and (\ref{eq:mass}). We calculate integrals numerically using the composite Simpson method. Then the specific AM is computed as $j_{*}=J_{*}/M_{*}$. The error in $j_{*}$ is estimated as

\begin{equation}
\delta j_{*} = R_{d}\sqrt{\frac{1}{N}\sum_{i=1}^{N} \delta v_{i}^{2}+\left(\frac{v_{f}}{tan(i)}\delta i\right)^{2}+\left(v_{f}\frac{\delta D}{D}\right)^{2}},
\end{equation}

where $N$ is the number of data-points, $\delta v_{i}$ is the error in the velocity on each point of the rotation curve, $D$ is the distance to the galaxy, and $\delta D$ is the distance error. The error in $R_{d}$ is estimated as a $10\%$ \citep{2016AJ....152..157L}. Finally the error of $M_{*}$ is estimated as

\begin{equation}
\delta M_{*} = \sqrt{\left(\Upsilon_{*}\delta{L_{3.6}}\right)^{2}+\left(L_{3.6}\delta \Upsilon_{*} \right)^{2}+\left(2M_{*}\frac{\delta D}{D}\right)^{2}},
\end{equation}

where $\delta L_{3.6}$ and $\delta \Upsilon_{*}=0.13$ \citep[same used in][]{2019MNRAS.484.3267L} are the errors of the total luminosity at 3.6 $\mu m$ ($L_{3.6}$) and $\Upsilon_{*}$, respectively.

It is important to stress that the stellar specific AM of the SPARC sample has been measured multiple times by other authors using different methods. Thus, the purpose of measuring it again is not to provide new values, but to use them to make a direct comparison between LSBGs and HSBGs, and to later estimate their spins. The former procedure is basically the same as in \cite{2018A&A...612L...6P}, with the main difference being the filter criteria, which is more rigorous in their paper than ours. The reason we do not apply the same criteria is because for this work we need to have a significant number of LSBGs, which would be drastically reduced otherwise. Thus, better measurements of the AM of LSBGs might be required to allow for better results.

\subsection{Estimating the spin parameter}

If the spin parameter of the halo is to be measured, first we should decide if the original definition of spin has to be used, or if the definition given by \cite{Bullock_2001} in Equation (\ref{eq:spin}) is preferred. The main problem with the original definition is that it is not easy to accurately estimate the energy, because it depends on the density profile. For example, one way of measuring the spin is to follow  \citet{1998MNRAS.295..319M}, where the assumption is that the halo is an isothermal sphere, and from the virial theorem one finds that the energy can be expressed as

\begin{equation}
E=-\frac{GM_\textnormal{vir}^{2}}{2R_\textnormal{vir}}=-\frac{M_\textnormal{vir}V_\textnormal{vir}^{2}}{2}.
\end{equation}

Using a Navarro-Frenk-White profile (NFW; \citeauthor{1996ApJ...462..563N} \citeyear{1996ApJ...462..563N}), one finds that the energy is the same as that for an isothermal sphere, but multiplied by a factor of $F_{E}(c)$ that depends, in turn,  on the concentration factor $c$. To obtain the virial masses, in this work we use the results from \citet{2019ApJ...886L..11L}, in which they employ three different halo density profiles to estimate the halo mass, by fitting the SPARC rotation curves using the NFW profile, the Einasto profile \citep{Einasto_1965}, and the DC14 profile \citep{DC14}. For this reason, it is considered  that a better option would be to use the value $\lambda '$ of Equation (\ref{eq:spin}), instead of the definition given by Peebles \citep{Peebles_1971}, in order to avoid calculating the energy. Although, in fact, obtaining $\lambda '$ is actually equivalent of using $\lambda$, and assuming the halo is an isothermal sphere. Hereafter, $\lambda '$ will simply be referred as $\lambda$.

The virial radius can be obtained from the virial mass from the relation

\begin{equation}
M_\textnormal{vir}=\frac{4}{3}\pi R_\textnormal{vir}^{3}\Delta_{c}\rho_{c},
\end{equation}

where $\rho_{c}$ is the critical density of the Universe and $\Delta_{c}$ is the over-density constant. In order to be consistent with the halo mass results of \citet{2019ApJ...886L..11L}, we adopt $\Delta_{c}=200$ \citep[for a discussion about the different halo mass definitions, see][]{2001A&A...367...27W}, thus $M_\textnormal{vir}=M_{200}$ is redefined and $R_\textnormal{vir}=R_{200}$. Using this value, the virial radius looks like

\begin{equation}
R_{200}=\left(\frac{M_{200}G}{100H^{2}(z)}\right)^{1/3},
\end{equation}

with

\begin{equation}
H(z)=H_{0}\sqrt{\Omega_{\Lambda}+(1-\Omega_{\Lambda}-\Omega_{0})(1+z)^{2}+\Omega_{0}(1+z)^{3}},
\end{equation}

the Hubble parameter, $z$ the redshift, and $H_{0}=100h$ Mpc$^{-1}$ km/s. We assume a $\Lambda$CDM cosmology,  with $\Omega_{\Lambda}=0.692$, $\Omega_{0}=0.308$, and $h=0.678$. Galaxy redshifts were taken from the NASA/IPAC Extragalactic Database \footnote{The NASA/IPAC Extragalactic Database (NED)
is operated by the Jet Propulsion Laboratory, California Institute of Technology,
under contract with the National Aeronautics and Space Administration.}.

Finally, in order to estimate the specific AM of the halo of the galaxies, $j_{h}$, two formation models are used in combination with three halo mass models. The first one used is the so-called classic model. In this model, the specific AM of the disk is equal to that of the halo  \citep[e.g.,][]{Fall_1980,1998MNRAS.295..319M}. The second model is the so-called biased collapse model \citep{vandenBosch:1998bi,2012ApJS..203...17R,2018MNRAS.475..232P}, in which $f_{j}$ depends on $f_{*}$.

\section{Models} 

\subsection{Halo mass models}
To obtain the virial masses of the galaxies, we use values computed by \cite{2019ApJ...886L..11L}. These authors use the NFW, Einasto and DC14 profiles for the DM halos to fit the SPARC rotation curves, and obtain the halo masses. Next, we present the corresponding radial profiles of the mass models, as well as a brief description of the properties of some of them.

We start with the NFW density profile. This comes from a N-body simulation in which the radial distribution of a particular halo is given by

\begin{equation}
\rho(r)=\rho_{c}\frac{\rho_{0}}{(r/r_{s})(1+r/r_{s})^{2}},
\end{equation}

where $\rho_{0}$ and $r_{s}$ are free parameters. This profile, however, fails when applied to LSBGs due to what is known as the ``Cuspy halo problem", where the observed density in the inner regions is lower than in the model. This implies that the NFW profile tends to overpredict the mass in some galaxies, and clearly the matter density at the galaxy bulges. Additionally, this model does not fit the rotation curves of LSBGs as well as other profiles allow, like DC14 \citep{2017MNRAS.466.1648K}.

The next profile model that is used is the Einasto one, defined by

\begin{equation}
ln(\rho(r)/\rho_{-2})=-2n\left(\left(\frac{r}{r_{-2}}\right)^{n}-1\right),
\end{equation}

where $r_{-2}$ is the radius at which the density profile has a slope of 2, $\rho_{-2}$ the density at that radius, and $n$ is a parameter that describes the shape of the profile. 

Finally, the DC14 profile, is based on a hydrodynamic simulation that introduces a dependence on $f_{*}$, in such a way that baryons affect the shape of the DM distribution, which goes as

\begin{equation}
\rho(r)=\frac{\rho_{s}}{\left(\frac{r}{r_{s}}\right)^{\gamma}\left( 1 + \left(\frac{r}{r_{s}}\right)^{\alpha} \right)^{(\beta - \gamma)/\alpha}},
\end{equation}

with $\rho_{s}$ and $r_{s}$ free parameters, $\alpha$, $\beta$, and $\gamma$ parameters depending on $f_{*}$. In particular, this model is worth considering since LSBGs should have smaller values of $f_{*}$ when compared to HSBGs, because of the low star formation rates and the dominance of DM in LGBGs.

\subsection{Classic disk formation model}

For the computation of the specific momentum of the halo, we consider two models. The first one, the classic disk formation (CDF hereafter) model, assumes that the specific momentum of the disk is equal to the halo specific AM, i.e. $j_{d}=j_{h}$. This assumption was first used by \cite{Fall_1980}, under the argument that baryons and DM experience the same external torques, before separating into two different components. This assumption also implies that $f_{j}$ is constant and close to $1$ for all galaxies, which is unlikely since $f_{j}$ depends on factors such as the dynamical friction, and should be less than $1$. But this is not the only problem. It turns out that even a constant $f_{j}<1$ does not work in this case. As shown by \citet{2018MNRAS.475..232P}, there is no constant $f_{j}$ that is able to reproduce the observed Fall relation. Keeping this in mind, the results derived from this model should, nevertheless, at least provide a limiting case for the values of the spin of the galaxies.

To calculate the specific AM of the disk, we chose the same procedure used to compute $j_{*}$, but considering only the disk component in Equation (\ref{eq:mom}).

\subsection{Biased collapse model}

Since the CDF model does not work for all cases, we also consider the biased collapse model \citep{vandenBosch:1998bi,2012ApJS..203...17R,2018MNRAS.475..232P}. In this model, $f_{j}$ is correlated with the star formation efficiency ($f_{*}$); and since $f_{*}$ is not constant, $f_{j}$ is not either. This model postulates that stars are formed, first, in the inner parts of the galaxy, where the gas density is higher and where the cooling is more effective. While the outer parts, with richer AM, fail to form stars. Meaning $j_{*}$ will be lower than the total specific AM. If we add that the momentum inside a radius $r$ grows as a power law with the gas mass, $j_\textnormal{gas}(<r)\propto M_\textnormal{gas}(<r)^{s}$ \citep{vandenBosch:1998bi,Bullock_2001,Dutton_2012}, this implies that

\begin{equation}
f_{j}=\left(\frac{f_{*}}{f_{b}}\right)^{s},
\end{equation}

where $f_{b}=0.157$ is the baryon fraction. Since $M_{*}$ and $M_{200}$  are available, $f_{j}$ can be calculated using this equation, obtaining $j_{h}$, which can later be used to get the spin. We use the exponent $s = 0.4$, as in \citet{2018A&A...612L...6P}.

The first consequence of considering this model applied to LSBGs and HSBGs, is that we expect $f_{*}/f_{b}$ to be smaller in LSBGs, and thus $f_{j}$ would be smaller and $j_{h}$ would be higher. Therefore, we should find higher spins for the LSBGs in this model when compared to the CDF model.

However, when considering this method, it is important to be aware of the glaring disadvantage that comes from the use of an exponent taken from fittings of this same sample. The model is basically forced to work, which makes the results seem artificial. On the other hand, this does not mean that the concepts and fundamentals of the model are wrong. It is worth considering both formation models, because the CDF model is known to not be accurate, and the biased collapse model seems too convenient. In other words, the truth can probably be found somewhere in between the results of the two formation models.

\section{Results} 

\subsection{Size, rotation and mass relations}

\begin{figure}[ht!]
\plotone{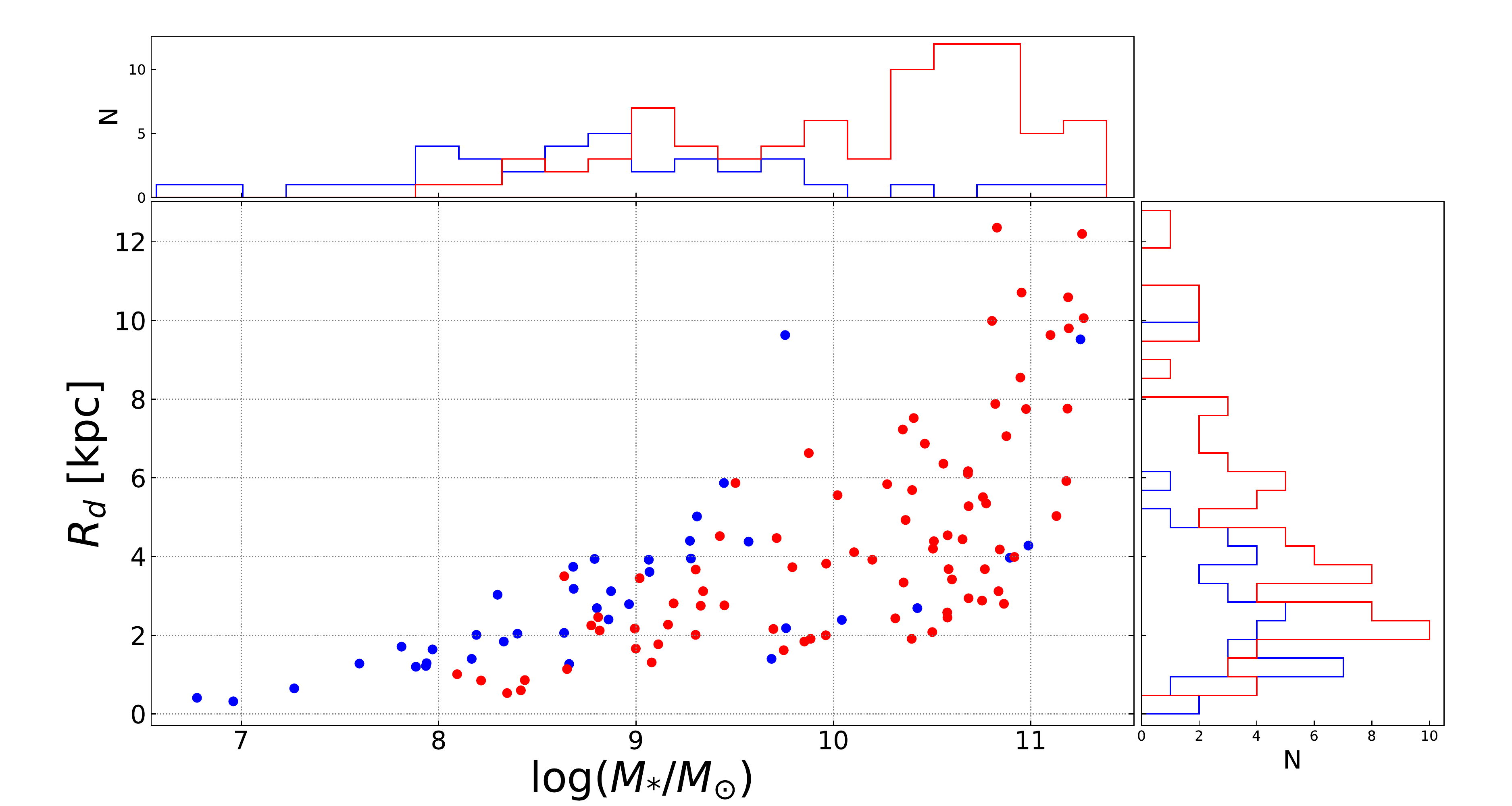}
\caption{Size-mass relation and distributions of LSBGs (blue) and HSBGs (red). It is apparent from this plot that the fraction of LSBGs is very low at masses higher than $log(M_{*}/M_{\odot})\sim 10$, and only few have smaller sizes compared to the HSBGs at the same stellar mass. \label{fig:sizemass}}
\end{figure}

\begin{figure}[ht!]
\plotone{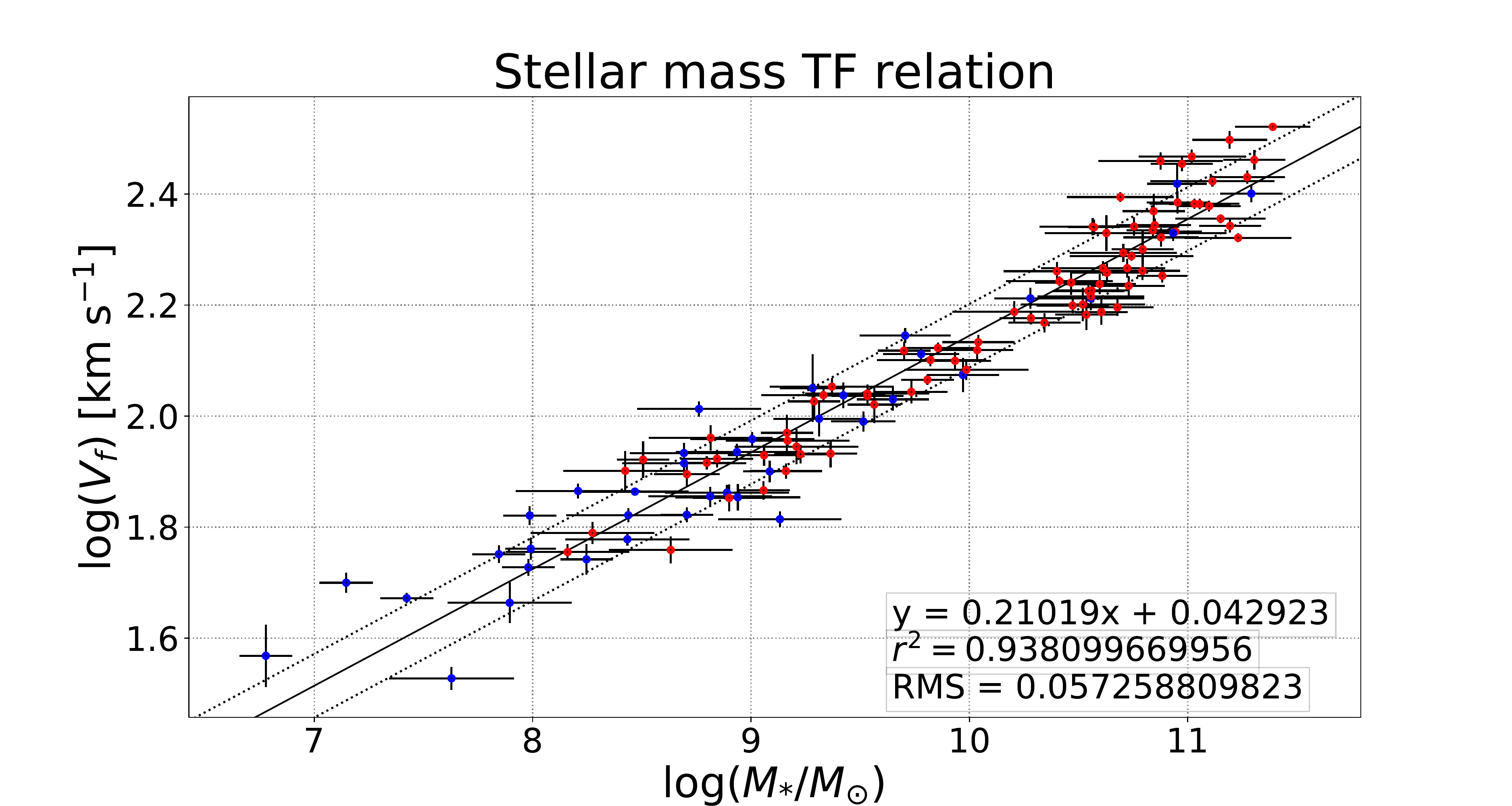}
\caption{Stellar mass Tully-Fisher relation. This is a replot of the results in \cite{10.1093/mnras/stz205}, but emphasizing the two populations of galaxies to show that both LSBGs (blue) and HSBGs (red) follow the same relation.}\label{fig:TF}
\end{figure}

Before going straight to the results of the AM, it is useful to first analyze the basic quantities involved in the computation of the AM. These are the size ($R_{d}$), rotation velocity ($v_{f}$), and mass ($M_{*}$) of the galaxies. To start, we present in Figure \ref{fig:sizemass} a size-mass relation together with their distributions. Figure \ref{fig:TF} shows the rotation velocity component and the stellar mass Tully-Fisher relation.

From Figure \ref{fig:sizemass}, it can be noted that LSBGs have a wide range of masses between $log(M_{*}/M_{\odot})\sim 7-12$, while HSBGs lie between $log(M_{*}/M_{\odot})\sim 8-12$. Note also that the number of high mass ($log(M_{*}/M_{\odot})> 10$) LSBGs is significantly lower than that of the HSBGs. This is important to note, since the AM increases with mass, although this does not really affect the trends in the $j-M$ diagram. It is also worth noting that low stellar mass LSBGs have larger sizes in general, but above $log(M_{*}/M_{\odot})\sim 10$, LSBGs are smaller than HSBGs.

As expected from previous measurements of the TF relation in the SPARC sample, all the galaxies follow the same TF relation, with LSBGs and HSBGs equally distributed around the linear fit of \ref{fig:TF}, which is only a replot of the results found in \citet{10.1093/mnras/stz205} but with different colors for LSBGs and HSBGs to better show that there is not a distinction between both populations in the TF diagram. It is worth noting that the TF relation is quite similar to the Fall relation, since $j_{*}\propto{R_{d}v_{f}}$. This means that should the size of HSBGs be the same as LSBGs, then both would follow the same Fall relation. However, Figure \ref{fig:sizemass} implies that this should not be expected to be the case. In addition, if we take into account the work of \citet{Dalcanton_1997}, galaxies with high spin values have baryons spread to a higher radius. Assuming the hypothesis that LSBGs have higher spin, and expecting the AM of baryons to be conserved during collapse, a higher specific AM in LSBGs could be reflecting this increase in radius.

\subsection{Fall relation for LSBGs and HSBGs}

\begin{figure}[ht!]
\plotone{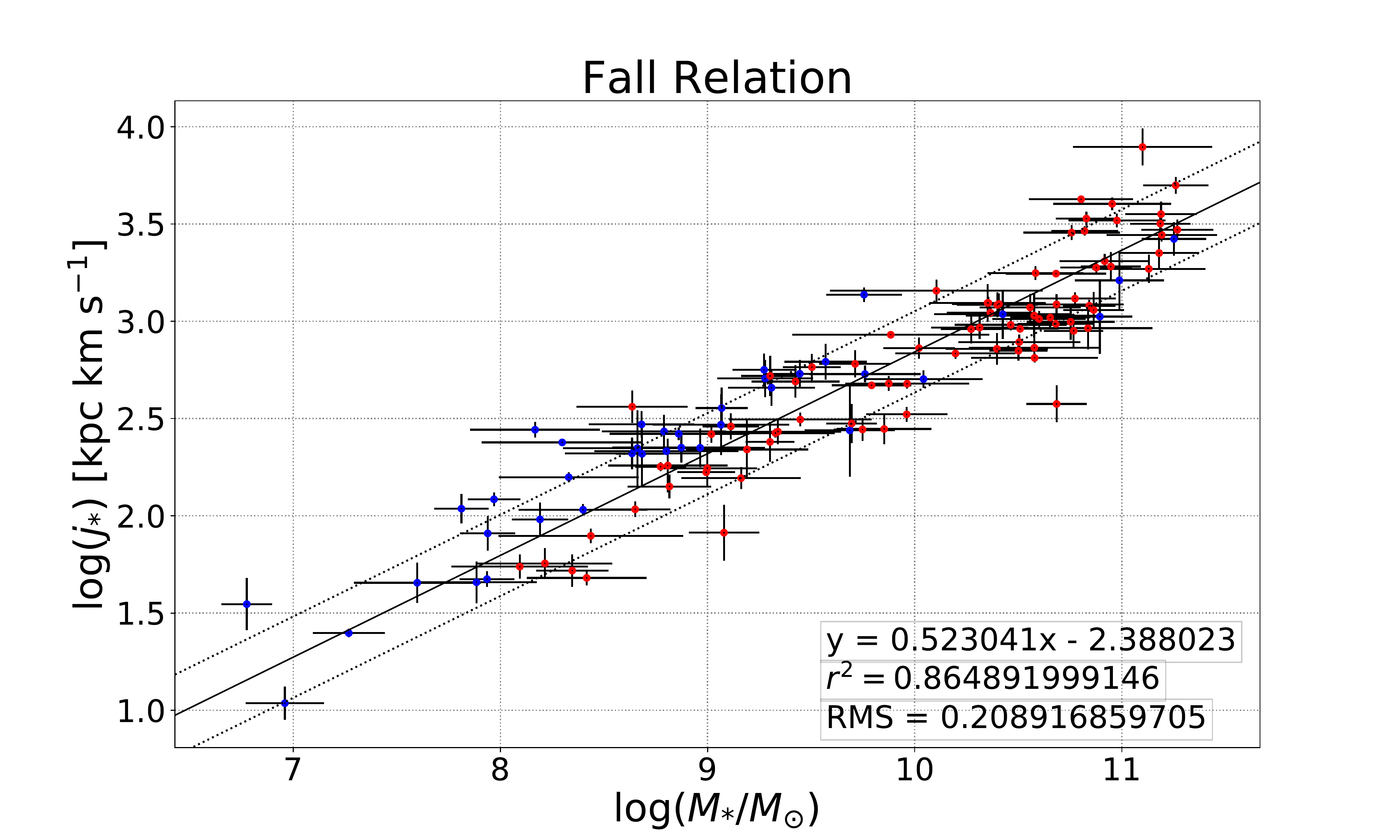}
\caption{Linear fit to the $j_{*}-M_{*}$ relation of the entire sample, with LSBGs (blue) and HSBGs (red). It is noted that LSBGs are mostly over the HSBGs, meaning that at a fixed stellar mass, they show, in general, higher stellar AM\label{fig:FALL}.}
\end{figure}

After measuring $j_{*}$ and $M_{*}$, we perform a linear fit of $log(j_{*})$ and $log(M_{*})$ of the entire sample of LSBGs and HSBGs in Figure \ref{fig:FALL} using the orthogonal distance regression method \citep{boggs_1990}. The corresponding error bars are calculated from the use of this method, together with the previously discussed uncertainties of the specific AM and stellar mass. This plot is similar to that of \cite{2018A&A...612L...6P}, but with a slightly larger sample (this one has more LSBGs included) due to fewer filter criteria. From Figure \ref{fig:FALL} it is clear that LSBGs have higher $j_{*}$ in comparison with HSBGs at a fixed $M_{*}$, indicating us that the position of galaxies in the Fall relation is strongly related with their surface brightness. However, it is worth to noting that the few massive LSBGs in the sample are not necessarily above the HSBGs, and in fact seem to be a bit lower. A possible explanation for this could be that these particular galaxies have earlier types than the rest of the LSBGs, and therefore one can speculate that during their evolution they might have been involved in more gravitational interactions lowering their AM. Though this is just a rough hypothesis, which is hard to confirm with only a small number of LSBGs in this mass range. An alternative explanation for this is presented at the end of the next section, where we estimate the spin and try to find a connection with the DM component. But perhaps we are in need of more and better data to have a complete and accurate picture. If we instead repeat this plot using $j_{*} \approx 2R_{d}v_{f}$ the massive LSBGs do show higher AM than HSBGs, which raises the question of whether we really have accurate measurements of $j_{*}$. More observations of massive LSBGs would definitely help in this regard.

\begin{figure}[ht!]
\plotone{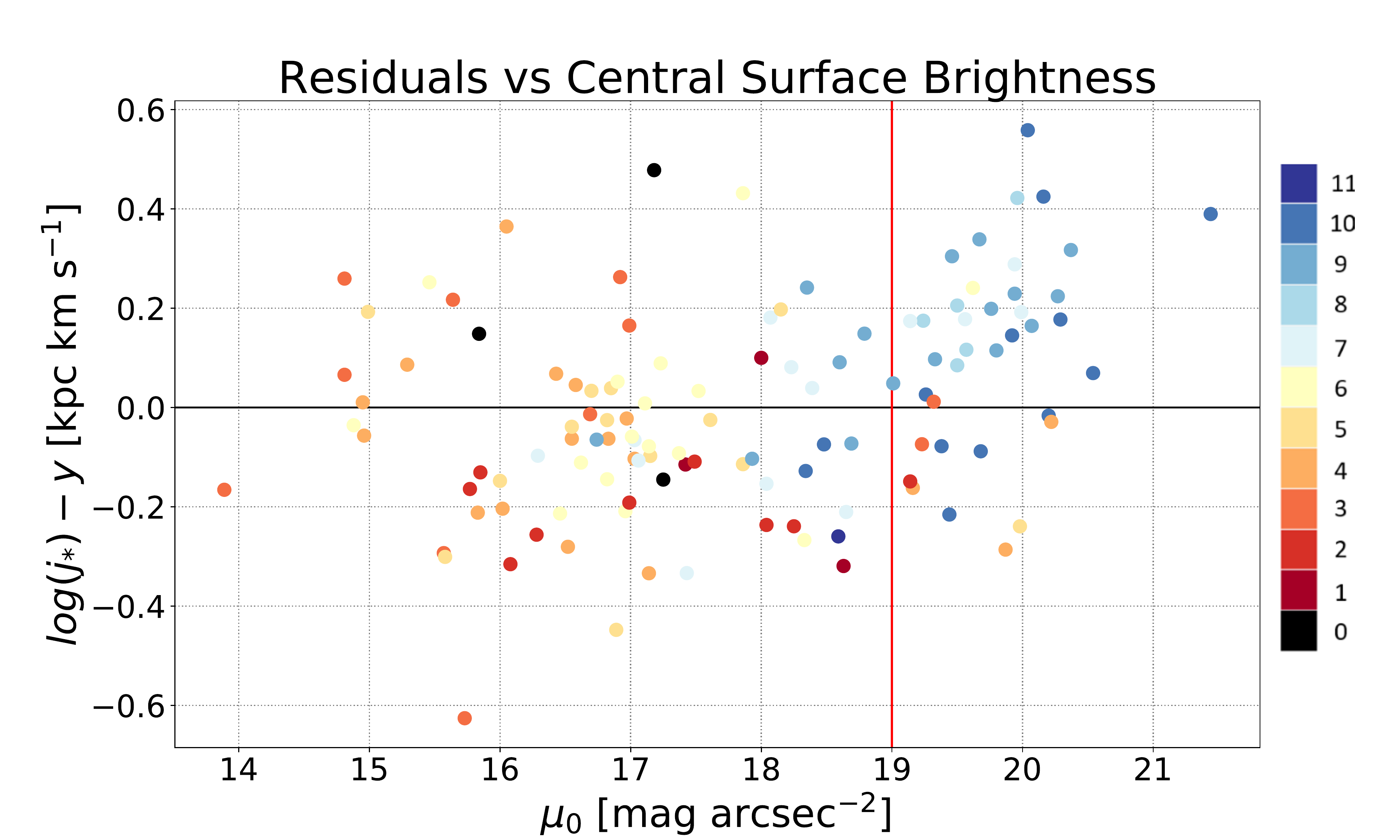}
\caption{Plot of the residuals from the linear fit of Figure \ref{fig:FALL} with central SB. Morphology is color coded following the numerical Hubble type of Figure \ref{fig:morph}. The red vertical line marks the separation between HSBGs and LSBGs.} \label{fig:ResidualvsSB}.
\end{figure}

Since most of the LSBGs are classified as late types and the Fall relation was first studied using the morphological type as a type or population discriminator, it may be challenged if these results are only a consequence of the morphology distribution of the sample. Hence, we can ask if differences stand if only one galaxy type is considered. In an attempt to answer this question we plot in Figure \ref{fig:ResidualvsSB} the residuals from the linear fit of Figure \ref{fig:FALL} against central SB, but also against Hubble types. From here, it is apparent that most of the LSBGs are of later types than HSBGs, but with a closer inspection it seems that even if LSBGs and HSBGs are considered to be of the same morphological type, LSBGs still tend to be located above the HSBGs, which is clearer if we look at morphology types between 7 and 10. This seems to indicate that the scatter in the Fall relation might be better explained by a range in stellar density, rather than only a range in morphology. However, it is also true that this is not a large enough sample to claim this with confidence. In any case, even if it is a consequence of the morphologies in the sample, LSBGs definitely seem to have higher specific AM as compared to HSBGs, finding an average difference of $0.174$ dex between the residuals of LSBGs and HSBGs.

From these results questions emerge: in the first place, why LSBGs have a higher specific AM than HSBGs? And, in which step of their evolution do they gain this systematically higher AM? One explanation could be mergers, as is usually invoked in the spiral versus elliptical problem. In this case, however, it seems unlikely due to the nature of LSBGs: considering the fact that LSBGs populate less dense environments (\citeauthor{2020MNRAS.tmp.3130P} \citeyear{2020MNRAS.tmp.3130P}, and references therein), we expect LSBGs to participate in fewer galaxy-galaxy interactions than HSBGs. On the other hand, one could suggest, based on the simulation from \cite{Dalcanton_1997}, that it is the spin of LSBGs that is higher than that of HSBGs, which could explain the higher $j_{*}$ found in the Fall relation.

Another possibility that is worth considering is that $f_{j} f_{*}^{-2/3}$ could be higher for LSBGs than for HSBGs. The last option of expecting $f_{*}$ to be smaller for LSBGs, does seem possible as well, but the factor $f_{j}$ also needs to be taken into consideration. Under the CDF model, $f_{j}$ should not be too different between LSBGs and HSBGs, meaning that a small $f_{*}$ might be sufficient to explain the higher $j_{*}$ in LSBGs. But, on the other hand, with the biased collapse model, small values of $f_{j}$ are expected for LSBGs, which might cancel-out the contribution that a small $f_{*}$ gives to the specific AM. If this last case is true, then higher spins are perhaps a better explanation for higher specific AM in LSBGs.

\subsection{Spin parameter for LSBGs and HSBGs}

Before looking at the spin distribution for galaxies, we can look at values in Table \ref{tab:1}, containing the mean fractions of $f_{*}$ and $f_{j}$ for LSBGs and HSBGs, and for every different model. We see that $f_{*}$ is smaller for LSBGs in all mass models, and in the biased collapse model, $f_{j}$ is also smaller for the LSBGs. Therefore, we should expect $j_{h}$ to be higher for LSBGs. 

\begin{deluxetable*}{ccccccc}
\tablenum{1}
\tablecaption{Mean values of $f_{*}$ and $f_{j} $\label{tab:1}}
\tablewidth{0pt}
\tablehead{
\colhead{Ratio} & \colhead{CDF-NFW} & \colhead{CDF-Einasto} & \colhead{CDF-DC14} & \colhead{BC-NFW} & \colhead{BC-Einasto} & \colhead{BC-DC14}
}
\startdata
$\bar{f}_{*,\textnormal{LSB}}$ & $0.0112\pm0.0019$ & $0.0170\pm0.0033$ & $0.0110\pm0.0019$ & $0.0112\pm0.0019$ & $0.0170\pm0.0033$ & $0.0110\pm0.0019$ \\
$\bar{f}_{*,\textnormal{HSB}}$ & $0.0346\pm0.0033$ & $0.056\pm0.006$ & $0.0305\pm0.0031$ & $0.0346\pm0.0033$ & $0.056\pm0.006$ & $0.0305\pm0.0031$ \\
$\bar{f}_{j,\textnormal{LSB}}$ & $0.98\pm0.05$ & $0.98\pm0.05$ & $0.98\pm0.05$ & $0.298\pm0.014$ & $0.343\pm0.017$ & $0.299\pm0.014$ \\
$\bar{f}_{j,\textnormal{HSB}}$ & $0.946\pm0.022$ & $0.946\pm0.022$ & $0.946\pm0.022$ & $0.484\pm0.013$ & $0.559\pm0.016$ & $0.461\pm0.015$ \\
\enddata
\tablecomments{This table contains mean values and corresponding $f_{*}$ and $f_{j}$. BC stands for biased collapse. The values of $f_{*}$ are the same in both CDF and biased collapse models, because they only depend on the mass model. The fraction $f_{j}$ in the CDF model is equal to $j_{*}/j_{d}$ since within this model, the assumption is that $j_{d}=j_{h}$.}
\end{deluxetable*}

\begin{figure*}
\gridline{\fig{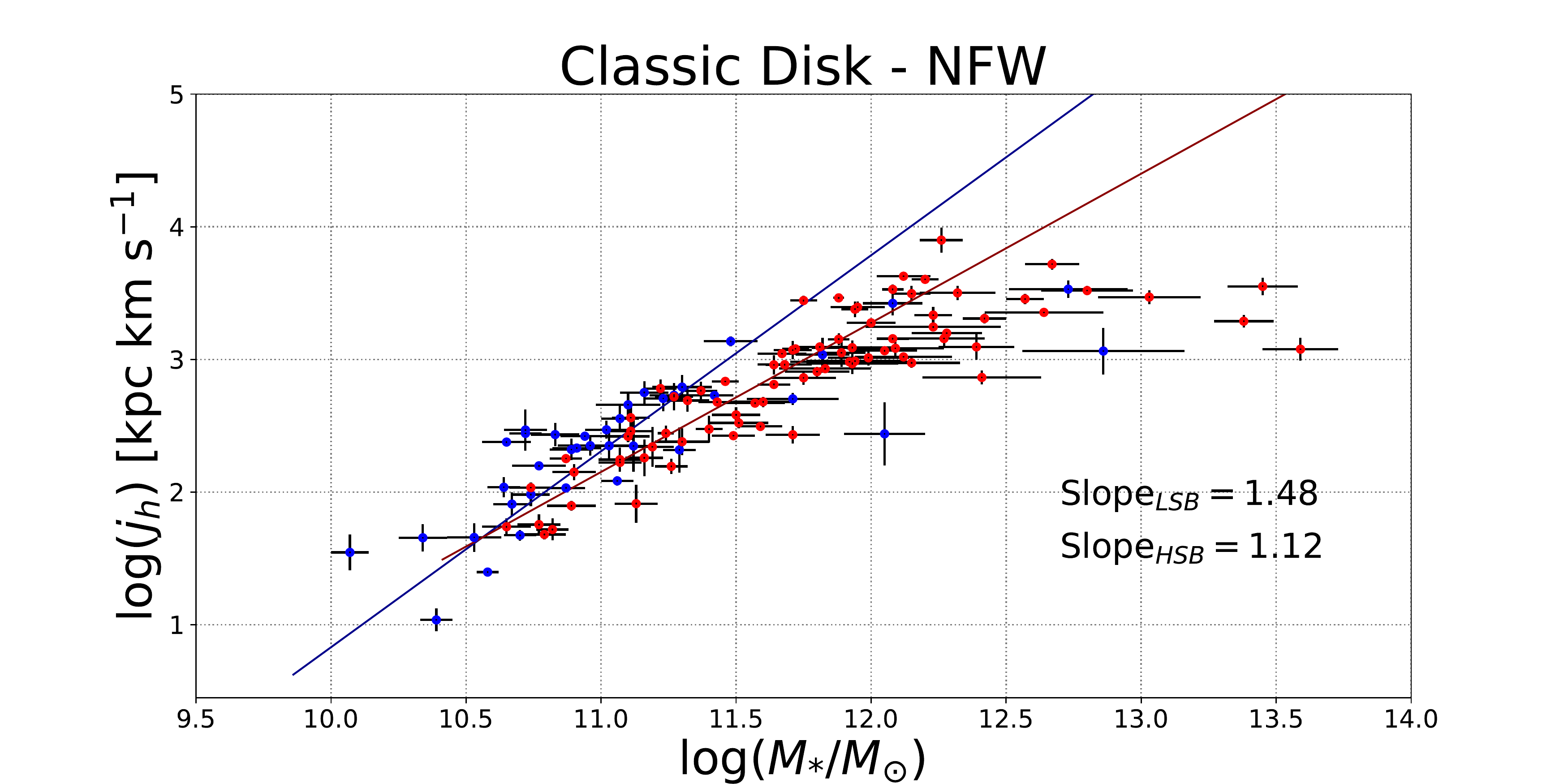}{0.38\textwidth}{}
          \fig{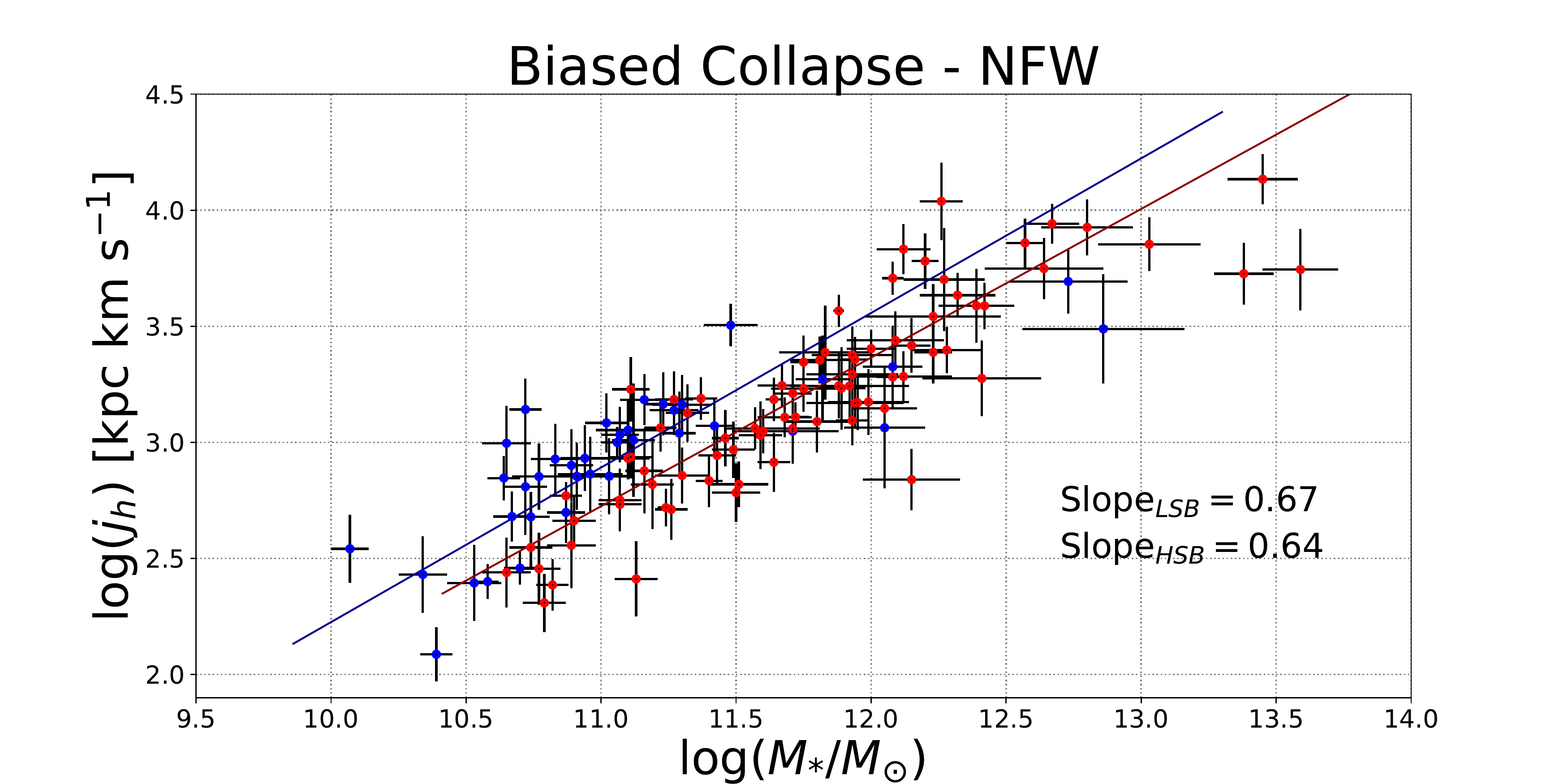}{0.38\textwidth}{}}
\gridline{\fig{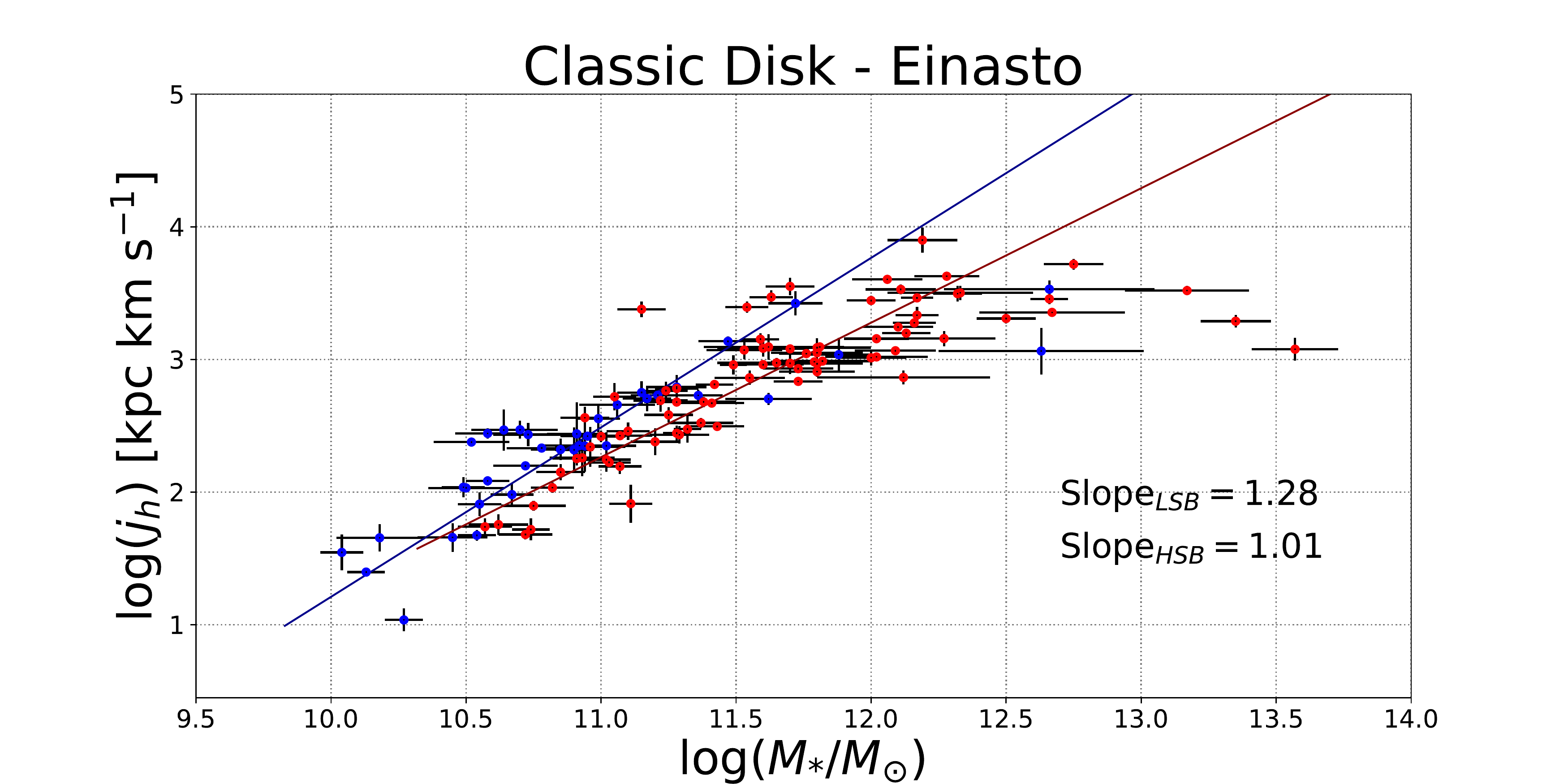}{0.38\textwidth}{}
          \fig{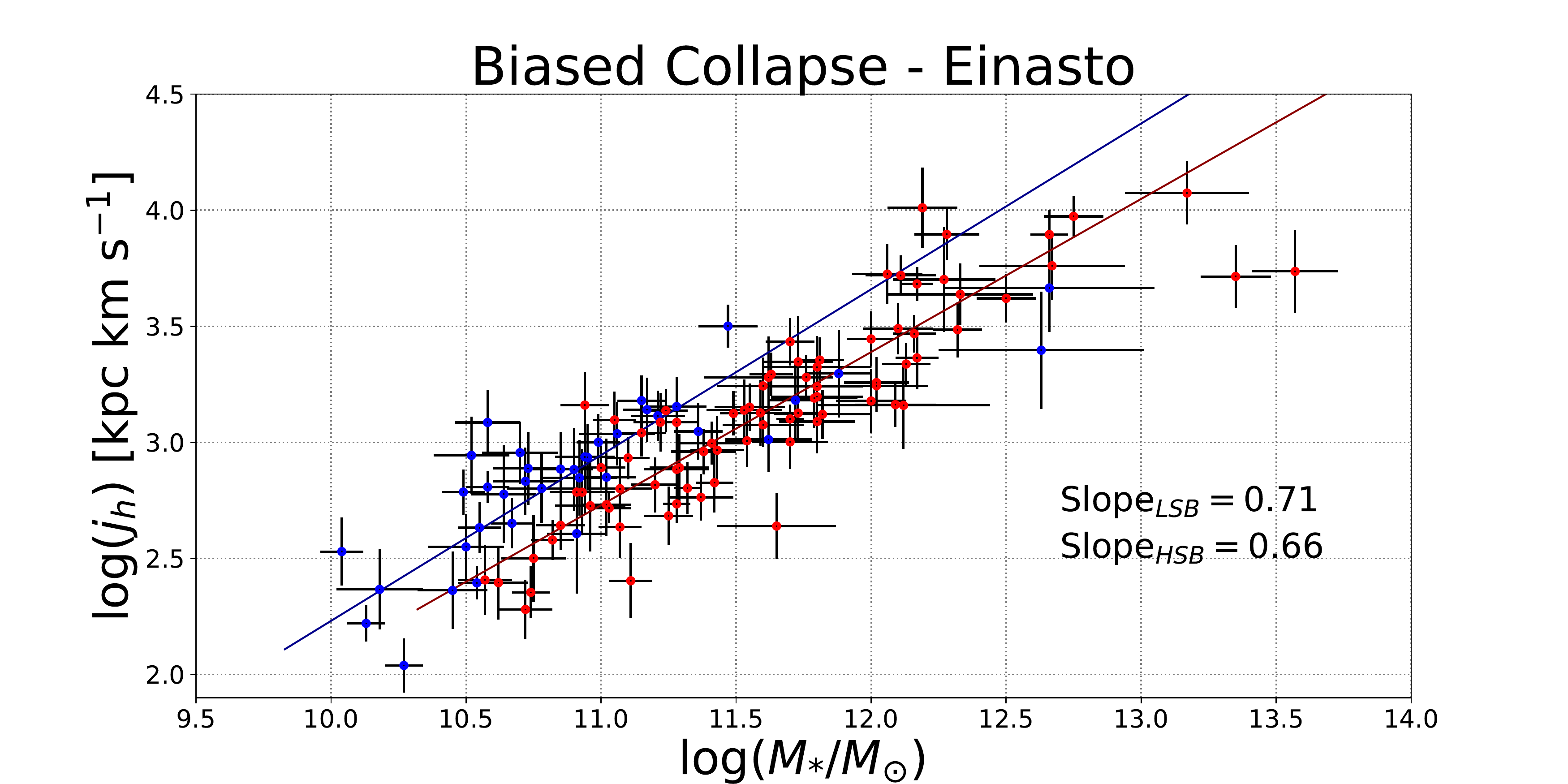}{0.38\textwidth}{}}
\gridline{\fig{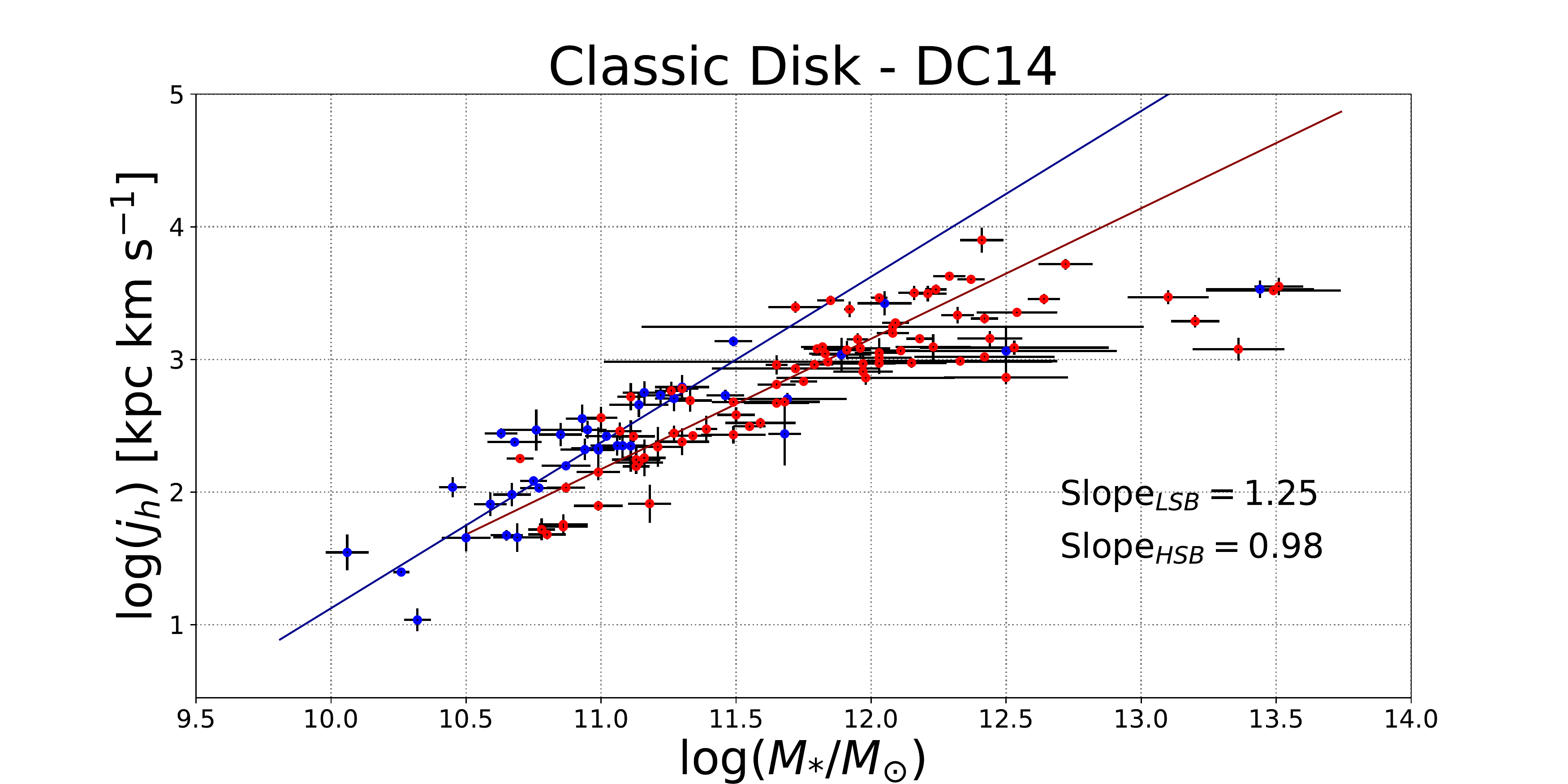}{0.38\textwidth}{}
          \fig{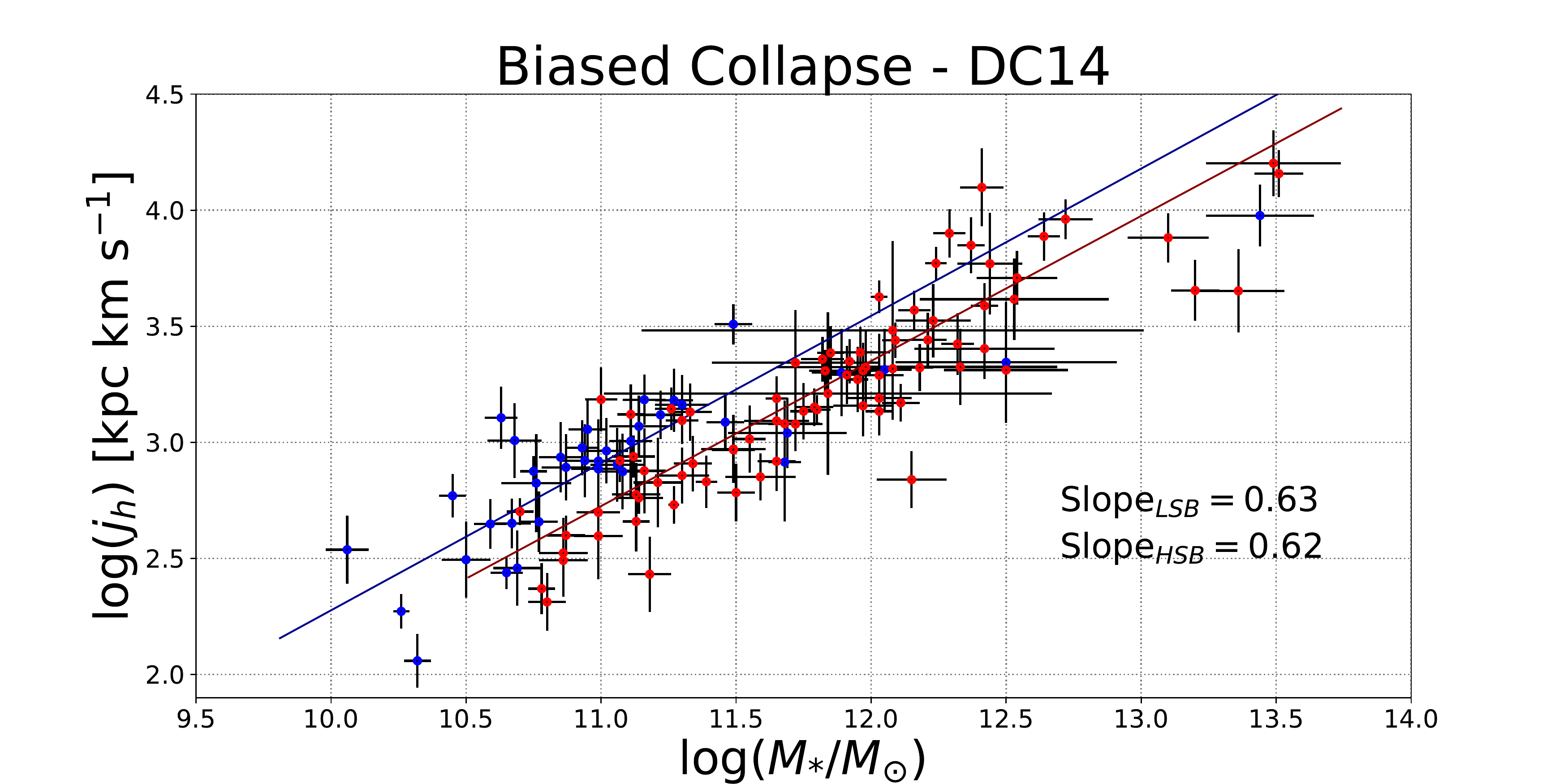}{0.38\textwidth}{}}
\caption{$j_{h}-M_{h}$ diagrams of the CDF and biased collapse models for LSBGs (blue) and HSBGs (red). From the CDF plots we find that, for both populations, and for all mass models, the slopes are higher than the value of $2/3$ that is predicted by equation (\ref{eq:jh}). There is also considerable scatter, with an rms scatter for LSBGs between 0.4-0.55, and 0.36-0.39 for HSBGs. On the other hand, if we consider the biased collapse model plots, we obtain slopes that are well fitted by the model by Equation (\ref{eq:jh}), both for LSBGs and HSBGs.    Also, the scatter is significantly lower, with a scatter of $0.23$ for the LSBGs in the NFW model.
\label{fig:JhMh}}
\end{figure*}

In order to compare models with observations, we carry out linear fittings of $j_{h}$ versus $M_{h}$. The fits explore whether or not we find slopes close to the $2/3$ value given by Equation (\ref{eq:jh}). In figure \ref{fig:JhMh} it is apparent that the trends of the diagrams in the CDF model do not follow the expected relation of Equation (\ref{eq:jh}), with the closest slope being $\sim1$ on the DC14 model for the HSBGs. Since the estimation of this result uses the values from the Fall relation, this means that this particular model does not match the observations, and it is inconsistent with relation (\ref{eq:jh}). What is found is that LSBGs are above HSBGs in Figure \ref{fig:JhMh}, but that is expected, given that we use in this model $f_{j}\sim 1$, and $M{*}$ with $M_{h}$ are related, so this basically can be traced back to the $M_{*}-M_{h}$ diagram. In fact, the large deviation from the trend at high masses is most likely due to the same deviation at high masses that occurs in the $M_{*}-M_{h}$ diagrams \citep[e.g., the diagrams in][] {2019ApJ...886L..11L}.

\begin{figure}[ht!]
\plotone{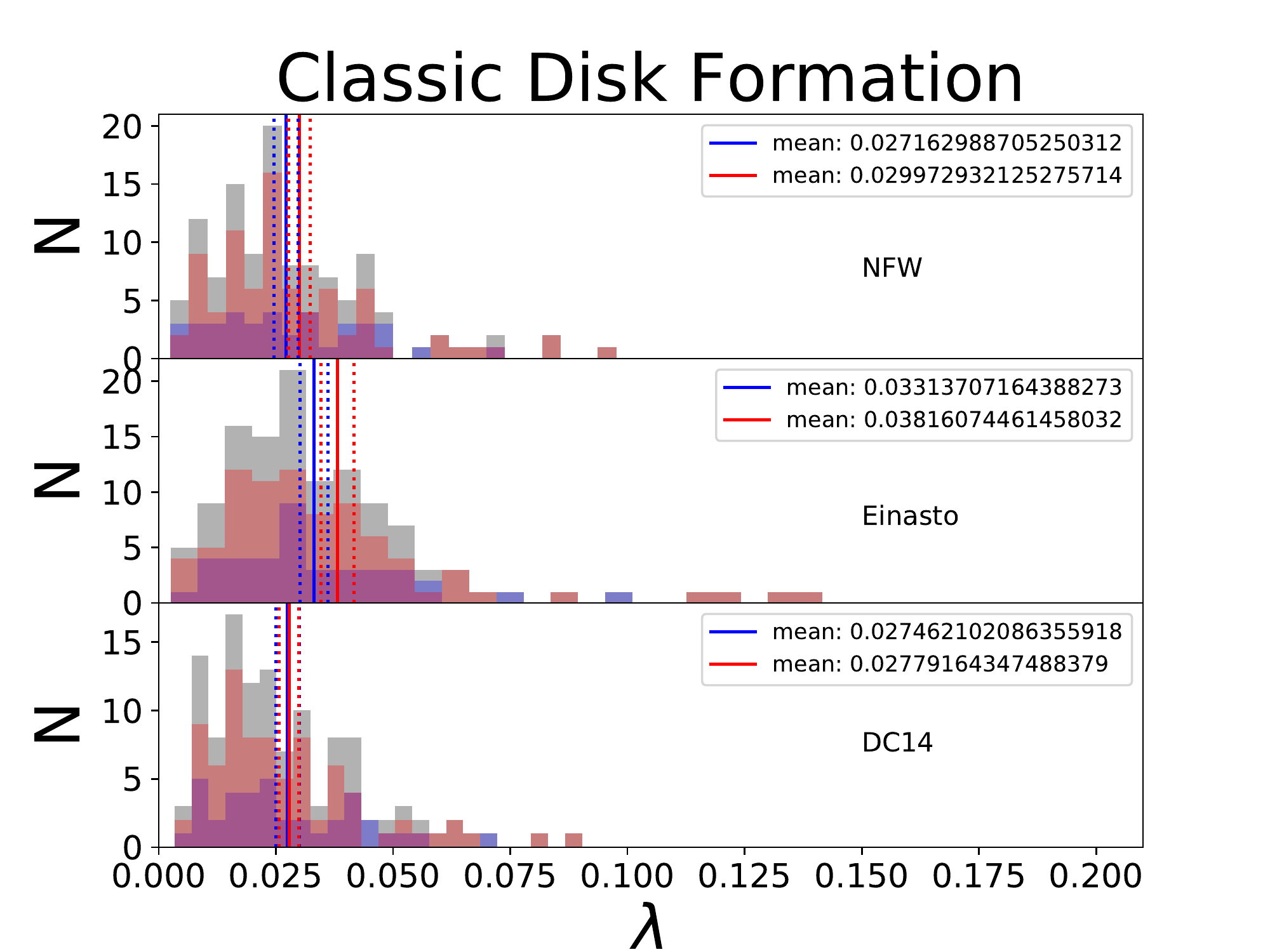}
\caption{Spin distributions of the CDF model for LSBGs (blue) and HSBGs (red). The vertical solid lines are the mean value of the distributions, and the dotted lines represent the error of the mean. From these plots we note that LSBGs always have an averaged spin that is slightly smaller than HSBGs. But, on the other hand, the averaged spin of the LSBGs is contained by the error bars of the HSBGs averaged spin, and vice versa. This suggests that there is not a big difference present in the spin of LSBGs and HSBGs for this model.\label{fig:CDFHist}}
\end{figure}

\begin{figure}[ht!]
\plotone{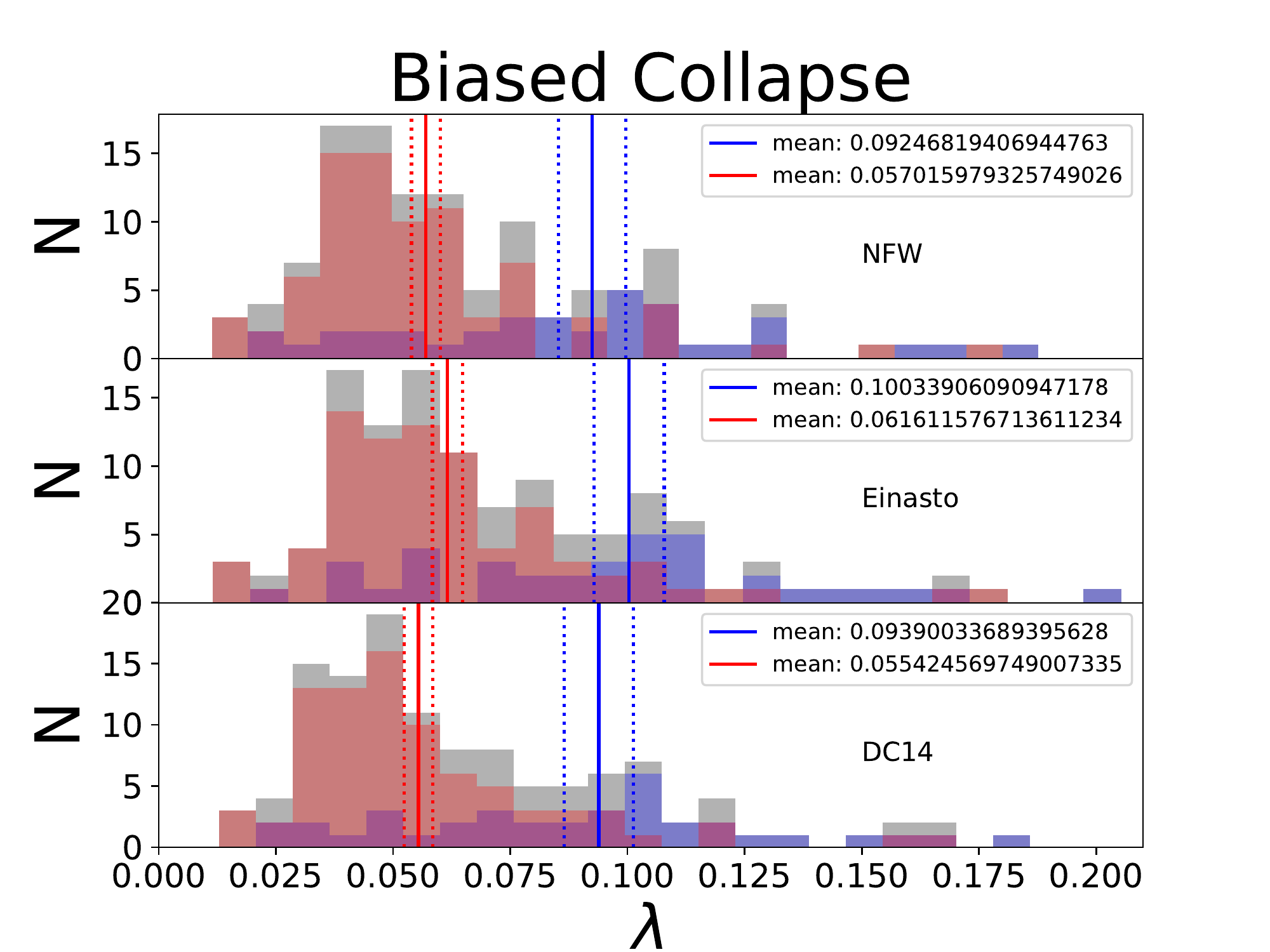}
\caption{Spin distributions of the biased collapse model for LSBGs (blue) and HSBGs (red). The lines represent the same values as in Figure \ref{fig:CDFHist}. Here results look very different from those in Figure \ref{fig:CDFHist}, with higher spin for LSBGs in all the mass models.\label{fig:BCHist}}
\end{figure}

When we look at the spin distributions in Figure \ref{fig:CDFHist} it becomes apparent that the LSBGs do not appear to have higher spin than HSBGs, as we came to expect from \cite{Dalcanton_1997}. On the contrary, the LSB population seems to have a slightly smaller average spin. In the NFW model, no indication whatsoever is found of a higher spin for LSBGs, but this might be because the NFW model does not fit the rotation curves well, specially in the case of LSBGs. Notwithstanding, in the case of the Einasto and DC14 models, however, it is shown that the peak of the LSBGs distribution might be at a slightly higher spin value. But, once again, the sample seems to be too small to assert this with enough confidence, as the peaks do not appear to be completely defined. This suggest that a larger sample would allow us to reveal that LSBGs have higher spin than HSBGs when using this model.

The results of the biased model, however, are quite different from the previous one. In fact, we found that the slope of the $j_{h}-M_{h}$ diagrams follows the theorical value of $2/3$ remarkably well, meaning that this model is a good candidate to explain the observations. From these diagrams we find once more that LSBGs lie above HSBGs, but with slopes more parallel to each other, with lower dispersion, and without the flattening of the specific AM at high masses that occurs in the previous model.

Regarding the spin distributions of Figure \ref{fig:BCHist},  a clear difference between HSBGs and LSBGs is present, with the latter having a very high spin that almost doubles the spin of the HSBGs in all of the halo mass models. This is interesting, since a recent study made by \cite{2019arXiv191004298P}, finds that the mean spin of LSBGs is $1.3$ to a factor of $2$ larger than that for HSBGs.  Although they estimated the spin with the CDF model and a much larger sample as well, which could indicate the slightly shifted peaks in Figure \ref{fig:CDFHist}, the Einasto and DC14 profiles could be perfectly in agreement with this tendency observed in a larger sample. Another interesting aspect when analyzing the biased collapse models is that spins extend to larger values, specially for the LSBGs which now have a wider range of spins than HSBGs, contrary to what the CDF model predicts, with a narrower distribution as compared to the HSBGs.

Since $f_{*}$ is smaller for LSBGs, and the biased collapse model is giving more consistent results, it seems more likely that the retained AM fraction ($f_{j}$) of LSBGs is smaller than that of HSBGs. This means that LSBGs would be more affected by the biased collapse than their normal counterparts. A possible reason for this, and borrowing again from the results in \cite{Dalcanton_1997}, could be that the mass in LSBGs is more spread out than in HSBGs. On the other hand, what is also interesting is that even when they retain less specific AM, they also end up having higher stellar AM than HSBGs, which, under the biased collapse scenario, is best explained by LSBGs having significantly higher spin values, as the results are showing. This is making the biased collapse model consistent and in agreement with the results from \cite{Dalcanton_1997}. 

It is worth pointing out that when we compare only the nonmassive galaxies, LSBGs have a higher average spin than HSBGs, including the CDF model. This means that the high mass part of the sample is responsible for the smaller spins in LSBGs. And because we have a small number of massive LSBGs to compare against massive HSBGs, we will need to include more massive LSBGs if we are to paint a complete picture. Another interesting point is the possible connection of these results with those in the Fall relation of Figure \ref{fig:FALL}, where both galaxy populations have a smaller slope than $2/3$, with LSBGs having a slightly smaller one than HSBGs, because of the drop in $j_{*}$ at high masses. If the biased collapse model is in play, it is possible that when the mass of the LSBGs is higher, they are finally more affected by the biased collapse, due to an expected increase in radius and a decreasing density that lowers their star formation efficiency, and thus, the specific AM consequently becomes smaller.

\section{Summary and Conclusion}

Using a sample taken from the SPARC database, we find that LSBGs have a higher stellar AM than HSBGs, locating themselves higher in the Fall diagram, with an average difference of about 0.174 dex higher than HSBGs. This is true even if we compare galaxies of the same morphological type. Additionally, we apply a combination of three different mass models, and two different formation models, to compute the spin parameter for a total of six different scenarios. Within the CDF model, where the stellar retained fraction is constant and close to 1, we find no clear-cut difference between the two populations. Only one unclear result, in two of the mass models, where the peak in the histogram of LSBGs seems to position more to the right than the peak of HSBGs, is evident. However, given that only 38 LSBGs were included, it is suggested that with a bigger sample, a clearer definition of the distribution peaks could be obtained, leading to LSBGs having higher spins in the CDF scenario. On the other hand, with the biased collapse model, which proved to be the most consistent model with the observed Fall relation, HSBGs and LSBGs have very different distributions, with LSBGs clearly having higher spin values. With the mean spin of LSBGs being about $\sim 2$ times higher than HSBGs in all the mass models.

The results in this work provide an observational --built-in-- comparison between the two populations in question, but there is still room for further research for a better understanding of the AM of LSBGs. The exact meaning in the values of the slopes for both HSBGs and LSBGs in the Fall relation still begs to be confirmed. Is the spin of LSBGs really higher than HSBGs? Is a higher spin in LSBGs the sole reason for a higher stellar AM when compared with HSBGs? Is the biased collapse affecting massive LSBGs differently? This paper presents some possible answers to these questions, but the full picture has yet to be revealed, and this is, in part, because some of the more general AM problems are not fully resolved.

On another note, subjects that are worth exploring in future studies could include measuring the gas contribution to the AM, which was not possible to attempt in this work due to the lack of enough H\,\textsc{i} mass density profiles for tracing gas distribution. This would broaden the understanding of the AM distribution in galaxies, and could also be used to make better estimations for the CDF model, by including the contribution of the gas into the specific momentum of the disk. Also, a larger, well measured sample, would be welcome, especially for massive LSBGs, since it should allow for clearer and more complete results of the spin distribution of the galaxies.

\acknowledgments
The authors acknowledge support from CONICYT project Basal AFB-170002. We acknowledge the anonymous referee for providing suggestions which helped to improve this paper

\bibliography{main}{}
\bibliographystyle{aasjournal}

\end{document}